\documentclass[preprint]{JHEP3} 

\usepackage{epsfig,multicol}

\newcommand\fverb{\setbox\pippobox=\hbox\bgroup\verb}
\newcommand\fverbdo{\egroup\medskip\noindent%
            \fbox{\unhbox\pippobox}\ }
\newcommand\fverbit{\egroup\item[\fbox{\unhbox\pippobox}]}
\newbox\pippobox

\newcommand{\be}{\begin{equation}}
\newcommand{\ee}{\end{equation}}
\newcommand{\ba}{\begin{eqnarray}}
\newcommand{\ea}{\end{eqnarray}}
\newcommand{\bt}{\begin{table}}
\newcommand{\et}{\end{table}}
\newcommand{\brt}{\begin{ruledtabular}}
\newcommand{\ert}{\end{ruledtabular}}
\newcommand{\btu}{\begin{tabular}}
\newcommand{\etu}{\end{tabular}}

\def\ov{\over}

\title{One-Loop Maximal Helicity Violating Amplitudes in $N=4$ Super Yang-Mills Theories}

\author{Mingxing Luo and Congkao Wen\\
Zhejiang Institute of Modern Physics, Department of Physics \\
Zhejiang University, Hangzhou, Zhejiang 310027, P R China \\
    E-mail: \email{luo@zimp.zju.edu.cn, wenagua@sina.com}}
\date{\today}        

\preprint{\hepth{0410045}}  

\abstract{One-loop maximal helicity violating (MHV) amplitudes in
$N=4$ super Yang-Mills (SYM) theories are analyzed,
using the prescription of Cachazo, Svrcek, and Witten (CSW). 
The relations between leading $N_c$ amplitudes $A_{n;1}$ 
and sub-leading amplitudes $A_{n;c}$ obtained by the CSW prescription
are found to be identical to those obtained from conventional field theory calculations. 
Combining with existing results, this establishes the validity of the CSW prescription
to one-loop in the calculation of MHV amplitudes in $N=4$ SYM theories of finite $N_c$.}

\keywords{MHV, One-loop, Supersymmetry, Yang-Mills Theory}

\begin{document}


\section{Introduction}

In perturbative quantum field theories, 
the Feynman diagram technique provides a standard method for calculating scattering amplitudes, 
but not necessary an efficient one. 
In practice, the method could be tedious and may require huge amount of calculations. 
On the other hand, the end results of most calculations are much simpler than the intermediate steps.
The simplification can be drastic if the theory is constrained by symmetries, 
though the calculation could instead be involved due to the proliferation of particles and couplings. 
For example, the Parke-Taylor formula for
maximal helicity violating (MHV) amplitudes in Yang-Mills theories
can be put within a single line which summarizes enormous number of
Feynman diagrams \cite{PT}. 
The simplification goes further if the theory is constrained by extra symmetries. 
For $N=4$ supersymmetric Yang-Mills (SYM) theories, 
one-loop results also assume extremely simple forms \cite{bern1,bern2}.

One is thus naturally led to search for new methods for specific
theories which would take the symmetries into account in the first place. 
In the specific case of Yang-Mills theories, 
various techniques were developed and tremendous progress has been achieved. 
Among them, the color decomposition method and the spinor helicity technique \cite{xzc} have been
proved to be extremely useful and efficient \cite{rvs}.

Recently, Witten pointed out a deep relation \cite{witten} between $N=4$ SYM
theories and one type B topological string theory, by re-expressing
SYM scattering amplitudes in the language of twistor theories \cite{penrose}.
Unfortunately, this relation breaks down beyond tree level,
as the supergravity part in the string construction does not decouple
form the SYM part \cite{wittenB}. 
However, taking advantage of insights thus gained 
and by a careful analysis of known helicity amplitudes, 
Cachazo, Svrcek, and Witten (CSW) \cite{csw} proposed a novel
prescription to calculate tree level amplitudes, 
which uses the MHV amplitudes as vertices to construct all other amplitudes. 
The efficiency of the method is phenomenal and the validity of the method has
been checked by various tree level calculations \cite{wuzhu,khoze,ggk,kosower,wuzhu2,bbk,wuzhu3,gk,zhu}. 
In \cite{bst}, the method has been extend to the calculation of one-loop MHV amplitudes and the
validity of the method is reconfirmed in the large $N_c$ limit.
The twistor-space structure of one-loop amplitudes are further studied in \cite{csw2,csw3}.
On the other hand, tree-level
amplitudes were also obtained from connected curves in twistor string theories \cite{rsv1,rsv2,rsv3}.

In this paper, one-loop MHV amplitudes in general $N=4$ SYM theories are calculated beyond the large $N_c$ limit,
by including all other diagrams which are not survived at the large $N_c$ limit. 
We reproduce the same relation between leading $N_c$ amplitudes $A_{n;1}$
and sub-leading amplitudes $A_{n;c}$ obtained by the CSW prescription
as those from conventional field theory calculations \cite{bern1}. 
Combining with results in \cite{bst}, this establishes the validity of the CSW approach
to one-loop in the calculation of MHV amplitudes in SYM theories, without taking the large $N_c$ limit.
The CSW prescription is motivated by a string construction and
the string construction so far seems mainly to be related to the large $N_c$ limit.
The sub-leading results obtained in this paper indicate a wider applicability.
This calls for further inquiry of the rationale behind the prescription.

The paper will be organized as follows. 
In section 2, we review one-loop results from conventional field theory
methods and the CSW prescription. 
In section 3, we analyze one-loop MHV amplitudes of four and five external particles
by using the CSW prescription. 
In section 4, the analysis is repeated for arbitrary number of external particles. We conclude in section 5.

\section{Review of existing one-loop results and the CSW prescription}
In four dimensional space-time, a momentum $k_\mu$ can be
expressed as a bispinor $k_{a\dot{a}}=k_\mu \sigma^\mu_{a\dot{a}}$. 
For massless particle, $k^2=0$, 
the momentum can be factorized $k_{a\dot{a}}=\lambda_a \tilde{\lambda}_{\dot{a}}$ 
in terms of spinors $\lambda_a$, $\tilde{\lambda}_{\dot{a}}$ 
of positive and negative chirality. 
Spinor products are defined to be
$\langle\lambda_1,\lambda_2\rangle=\epsilon_{ab}\lambda^a_1\lambda^b_2$
and $\langle\tilde{\lambda}_1,\tilde{\lambda}_2\rangle=
\epsilon_{\dot{a}\dot{b}}\tilde{\lambda}^{\dot{a}}_1\tilde{\lambda}^{\dot{b}}_2$,
which are usually abbreviated as $\langle 1,2 \rangle$ and
$[1,2]$.

At tree level, the scattering amplitudes of $n$ gluons with one or
none opposite type of helicity vanish. The amplitudes with two
negative helicity are called maximally helicity violating (MHV)
amplitudes. For $N=4$ SYM theories, a MHV amplitude is given
by the generalized Parke-Taylor formula which includes particles
of all helicity \cite{nair}: 
\be 
i (2\pi)^4 \delta^{(4)} \left( \sum_{i=1}^n \lambda_i \tilde{\lambda}_i \right) 
                 \delta^{(8)} \left( \sum_{i=1}^n \lambda_i \eta^i \right) 
                  A_n \left(\{k_i,\lambda_i,a_i\} \right) 
\ee 
where  $\eta_A^i$ are
anti-commuting variables, $A$ is an index of the anti-fundamental
representation of $SU(4)$; $k_i$, $\lambda_i$, and $a_i$ are the
momentum, helicity, and the color index of the $i$-th external
particles, respectively; 
\be A_n \left( \{k_i,\lambda_i,a_i\} \right) = 
      \sum_{\sigma \in S_n/Z_n} {\rm Tr} \left(T^{a_{\sigma(1)}} \cdots T^{a_{\sigma(n)}} \right) 
      \prod_{i=1}^n {1 \ov \langle \sigma(i),\sigma(i+1)\rangle}
\ee 
where $S_n/Z_n$ is the set of non-cyclic permutation of
$\{1,\cdots, n\}$. The $U(N_c)$ generators $T^a$ are the set of
hermitian $N_c\times N_c$ matrices normalized such that ${\rm Tr}
(T^a T^b) = \delta^{ab}$.
Here and after the gauge coupling constant $g$ is not included, 
but can be easily recovered when needed. 
The supersymmetric amplitudes can be expended in powers of the $\eta_A^i$, 
and each term of this expression corresponds to a particular scattering amplitude.

To one-loop, the results take the form \cite{bern1}
\be 
A_n^{\rm 1-loop} \left( \{k_i,\lambda_i,a_i\} \right)
  = \sum_{c=1}^{[n/2]+1} \sum_{\sigma \in S_n/S_{n;c}} Gr_{n;c}(\sigma) A_{n;c}(\sigma) 
\label{eq2.3}
\ee 
where $S_n$ is the set of all permutation of $n$ objects 
and $S_{n;c}$ is the subset leaving $Gr_{n;c}$ invariant. 
The color factors are
\ba 
Gr_{n;1} &=& N_c {\rm Tr} \left( T^{a_1} \cdots T^{a_n} \right) \nonumber \\
Gr_{n;c} &=& {\rm Tr} \left( T^{a_1} \cdots T^{a_{c-1}} \right)
             {\rm Tr} \left( T^{a_c} \cdots T^{a_n} \right) 
\ea 
The simplest kinematic factors are proportional to the corresponding tree level ones \cite{bern1,bern2}
\be 
A_{n;1}(1,\cdots, n)=  V_n \prod_{i=1}^n {1 \ov \langle i,i+1\rangle} 
\label{eq2.5}
\ee 
where $V_n$ is an universal function dependent
of external momenta but not of the helicities. $A_{n;c>1}$ can be
obtained by performing appropriate sums over permutations of
$A_{n;1}$: 
\be 
A_{n;c}(1,\cdots, c-1; c,\cdots n) = (-)^{c-1} \sum_{\sigma \in COP \{\alpha\}\{\beta\}} A_{n;1}(\sigma) 
\label{Anc_field}
\ee
where $\alpha_i \in \{\alpha\} \equiv \{c-1,\cdots,1\}$,
 $\beta_i \in \{\beta\} \equiv \{c,c+1,\cdots,n\}$ and $COP
 \{\alpha\}\{\beta\}$ is the set of all permutations of
 $\{1,\cdots,n\}$ with $n$ held fixed that preserve the cyclic
 ordering of $\alpha_i$ within $\{\alpha\}$ and of $\beta_i$ within
 $\{\beta\}$, while allowing for all possible relative ordering
 of $\alpha_i$ with respect to $\beta_i$.

In \cite{csw}, it was proposed that these MHV amplitudes can be
continued to off-shell and be used as vertices to generate tree diagrams for all non-MHV amplitudes.
These new diagrams are dubbed as MHV diagrams. Including all relevant factors,
the $n$-points MHV vertex is 
\be 
\mathcal{V}_n = i {\rm Tr} \left( T^{a_1} \cdots T^{a_n} \right)(2\pi)^4 
        \delta^{(4)} \left( \sum_{i=1}^n \lambda_i \tilde{\lambda}_i \right) 
        \delta^{(8)} \left( \sum_{i=1}^n \lambda_i \eta^i \right)
        \prod_{i=1}^n {1 \ov \langle i,i+1 \rangle} 
\ee 
The propagator for internal lines of momentum $k$ is simply taken as $i/k^2$. The
issue is how to define the corresponding spinor for particles with
off-shell momentum $k$. The CSW prescription is to take an
arbitrary spinor of positive helicity $\tilde{\xi}$ and define $\lambda_a
= k_{a \dot{a}} \tilde{\xi}^{\dot{a}}.$

\begin{figure}[h]
\begin{center}
\leavevmode
{\epsfxsize=4.0truein \epsfbox{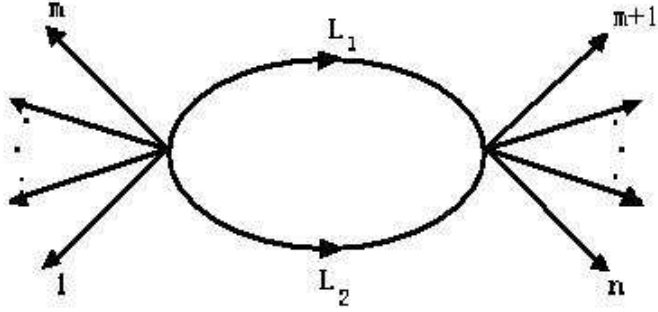}}
\end{center}
\caption{One leading order one-loop MHV diagram, using MHV amplitudes as interaction vertices,
with the CSW off-shell prescription.}
\end{figure}

Extending these rules to one loop and to the leading order of $N_c$, 
the correct expression of $A_{n;1}$ is reproduced \cite{bst}.
For example, the MHV diagram in Figure 1 gives 
\ba 
A&=&i (2\pi)^4 \delta^{(4)}(P_L+P_R)  
    \int {d^4L_1 \ov L_1^2} \int {d^4L_2 \ov L_2^2} \int d^4\eta d^4\eta^{'}
    \delta^{(4)} \left( L_2+L_1 +P_L \right) A_L A_R 
\ea 
where $P_L$ is the sum of momenta on the left of the diagram. 
Other factors are,  
\ba
A_L & = & {\rm Tr} \left(T^a T^{a_1}\cdots T^{a_m}T^b \right) \delta^{(8)}(\Theta^L) 
          {1 \over \langle l_2,1 \rangle \langle m, l_1\rangle \langle l_1,l_2 \rangle} 
           \prod_{i=1}^{m-1} {1\over \langle i,i+1 \rangle}  \\ 
A_R& =& {\rm Tr}\left(T^b T^{a_{m+1}}\cdots T^{a_n} T^a \right)\delta^{(8)}(\Theta^R) 
          {1 \over \langle -l_1,m+1 \rangle \langle n, -l_2\rangle \langle -l_2, -l_1 \rangle} 
           \prod_{i=m+1}^{n-1} {1\over \langle i,i+1 \rangle}
\nonumber 
\ea 
where
\ba 
\Theta^L_{A\alpha} & = & \sum_{i\in L} \eta_A^i
    \lambda_\alpha^i + \eta_A^{'} l_{2\alpha} + \eta_A l_{1\alpha} \nonumber \\
\Theta^R_{A\alpha} &=&\sum_{i\in R} \eta_A^i 
    \lambda_\alpha^i - \eta_A^{'} l_{2\alpha} - \eta_A l_{1\alpha} 
\ea
and up to irrelevant overall factors,
\ba
 l_{\alpha}  =  L_{\alpha\dot{\alpha}} \tilde{\xi}^{\dot{\alpha}}, \ \ \
 \tilde {l_{\dot{\alpha}}}  =  \xi^{\alpha} L_{\alpha\dot{\alpha}} \nonumber
\ea

Integrating out $\eta$ and $\eta^{'}$, the fermionic part gives an overall 
$\delta^{(8)} (\sum_i^n \eta^i \lambda^i)$, which will be ignored in the rest of the paper.
Summing up the color indices, we get the rest expression,
\be 
A = Gr_{n;1} L_n \prod_{i=1}^n {1\ov \langle i,i+1\rangle} 
\ee 
where 
\be 
L_n = \int {d^4L_1 \over L_1^2} \int {d^4L_2 \over L_2^2} \delta^{(4)}(L_2+L_1 +P_L)
     {\langle 1, n\rangle \langle l_1 l_2 \rangle \over \langle n,l_1\rangle \langle l_1,1\rangle} 
     {\langle m, m+1\rangle \langle l_1 l_2 \rangle \over \langle m,l_2\rangle \langle l_2,n\rangle}
\ee

After an extensive calculation, $L_n$ is shown to be independent of $\tilde{\xi}$.
Including all inequivalent diagrams and summing up corresponding $L_n$'s,
the exact expression $V_n$ in Eq (\ref{eq2.5}) is reproduced. 
That is, the leading contribution in Eq (\ref{eq2.3}) is recovered, 
and the validity of the CSW prescription established in the large $N_c$ limit.

We now classify the one-loop MHV diagrams according to their topologies.
One-loop MHV diagrams with all external lines outside of the circle, 
such as the one in Figure 1, will be referred to as leading $N_c$ MHV diagrams, 
since they give the $G_{n;1}$ color factors;
those with external lines both outside and inside the circle,
such as the one in Figure 8, as sub-leading MHV diagrams, 
since they give the $G_{n;c>1}$ color factors. 
(One-loop MHV diagrams with all external lines inside of the circle
are identical to leading MHV diagrams, up to possible overall signs.)
As indicated above, the former give the leading amplitudes $A_{n;1}$.
The latter should give the sub-leading amplitudes $A_{n;c>1}$ 
if the CSW prescription is independent of the large $N_c$ limit.
The proof of this assertion is the main focus of this paper.

\section{Special cases of four and five external particles}

In this section, we show that the CSW prescription gives 
the full one-loop MHV amplitudes in cases of four and five external particles.

We note that one cannot make a loop out of a single MHV vertex.
Such diagrams would include the following $\eta$ integration,
\be
\int d^4\eta^l\delta^{8}(\Theta).
\ee
However,
\be
\Theta=\sum_{i=1}^{n}\eta^{i}_{A}\lambda^{\alpha}_{i}+\eta^{l}_{A}l^{\alpha}
-\eta^{l}_{A}l^{\alpha}=\sum_{i=1}^{n}\eta^{i}_{A}\lambda^{\alpha}_{i}
\ee
which is independent of $\eta^{l}$. The $\eta$ integral thus yields a null result.
The scattering amplitudes of $n$ gluons with one or
none opposite type of helicity also vanish, as expected.
One-loop MHV diagrams are constructed from two and only two MHV vertices. 

\begin{figure}[h]
\begin{center}
\leavevmode
{\epsfxsize=4.0truein \epsfbox{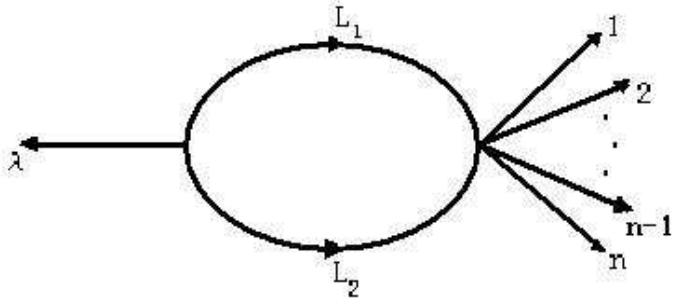}}
\end{center}
\caption{A one-loop MHV diagram with one three-leg MHV vertex, shown to give a null result.}
\end{figure}

Next we prove that one-loop MHV diagrams with three-leg MHV vertices also give null results.
We start with a simple diagram with all external lines outside of the circle,  as shown in Figure 2.
After integrating out the fermionic variables, this diagram gives 
\be
\int\frac{d^{4}L_{1}}{L_{1}^{2}}\frac{d^{4}L_{2}}{L_{2}^{2}}\delta^{4}(L_{1}+L_{2}+P_{\lambda})
\langle l_1,l_2 \rangle^2 F \label{eq3.3}
\ee
where $F$ is a complicated factor but irrelevant to our discussion.
Due to energy-momentum conservation, $L_{1}+L_{2}+P_{\lambda}=0$,
the off-shell spinors are related
\ba
(l_1)_\alpha & = &-(L_2)_{\alpha\dot{\alpha}} \tilde{\xi}^{\dot{\alpha}}
-\lambda_{\alpha\dot{\alpha}} \tilde{\xi}^{\dot{\alpha}} \nonumber \\
&=&-(l_2)_{\alpha}-\lambda_{\alpha\dot{\alpha}} \tilde{\xi}^{\dot{\alpha}}
\ea
up to irrelevant factors.
Since the integration in Eq. (\ref{eq3.3}) is independent of $\tilde{\xi}$ \cite{bst}, 
we can choose $\tilde{\xi}^{\dot{\alpha}}=\tilde{\lambda}^{\dot{\alpha}}$.
We have then $l_1=-l_2$ so the integrand in Eq. (\ref{eq3.3}) vanishes.
For general diagrams with external lines both inside and out side of the circle,
a theorem to be proved in next section states that they can be expressed as sums of
diagrams with all particles outside of the circle. This closes our proof.

Now we consider one-loop MHV diagrams of four external particles.
They are made of two MHV vertices and each vertex has four legs. 
The two diagrams in Figure 3a give the leading contribution with color factor 
$N_c {\rm Tr} \left( T^{a_1} T^{a_2} T^{a_3} T^{a_4} \right)$
\be
A_{4;1}(1,2,3,4)=
\int {d^4 L_1 \ov L_1^2} {d^4 L_2 \ov L_2^2} F_0(1,2,3,4) \label{eq3.5}
\ee
where 
\ba
F_0(1,2,3,4) & = & \delta^4 (L_1 + L_2+P_{1L}) h_1(1,2,3,4) +  \delta^4 (L_1 + L_2+P_{2L})  h_2(1,2,3,4)
\ea
with $P_{1L} = \lambda_1 \tilde{\lambda}_1 + \lambda_2 \tilde{\lambda}_2$,
 $P_{2L} = \lambda_1 \tilde{\lambda}_1 + \lambda_4 \tilde{\lambda}_4$ and
\ba 
h_1(1,2,3,4)=
\frac{1}{\langle 1,2\rangle\langle 2,3\rangle\langle 3,4\rangle\langle 4,1\rangle}
\frac{\langle 2,3\rangle\langle l_2,l_1\rangle}{\langle 2,l_1\rangle\langle l_1,3\rangle}
\frac{\langle 4,1\rangle\langle l_1,l_2\rangle}{\langle 4,l_2\rangle\langle l_2,1\rangle}
 \nonumber \\
h_{2}(1,2,3,4)=
\frac{1}{\langle 4,1\rangle\langle 1,2\rangle\langle 2,3\rangle\langle 3,4\rangle}
\frac{\langle 1,2\rangle\langle l_2,l_1\rangle}{\langle 1,l_1\rangle\langle l_1,2\rangle}
\frac{\langle 3,4\rangle\langle l_1,l_2\rangle}{\langle 3,l_2\rangle\langle l_2,4\rangle}
\ea
Eq (\ref{eq3.5}) gives the correct leading contribution.

\begin{figure}[h]
\begin{center}
\leavevmode
{\epsfxsize=5.0truein \epsfbox{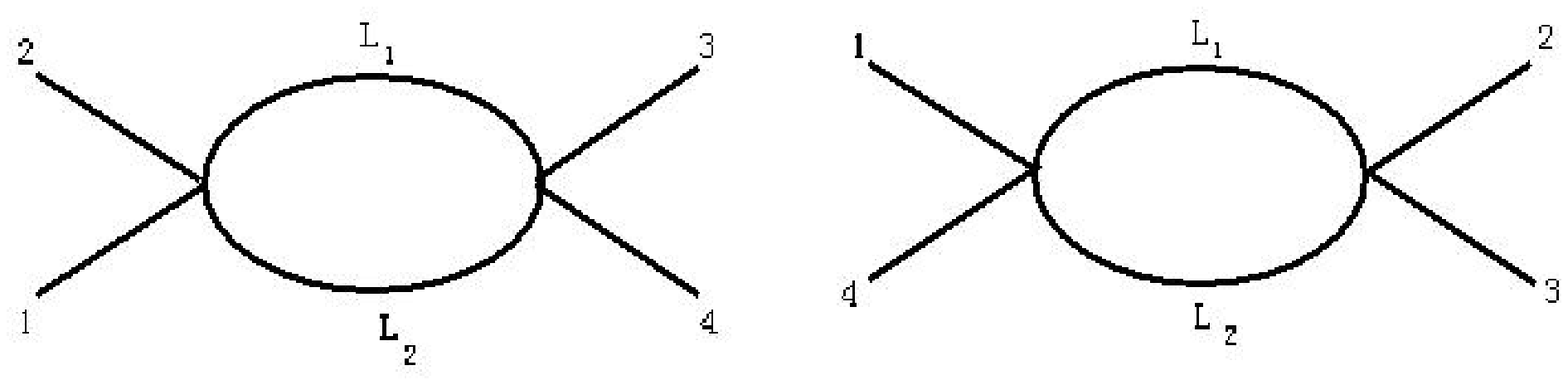}}
\end{center}
\caption{(a) Leading one-loop MHV diagrams for the special case of four external particles.}
\end{figure}
\setcounter{figure}{2}

\begin{figure}[h]
\begin{center}
\leavevmode
{\epsfxsize=2.50truein \epsfbox{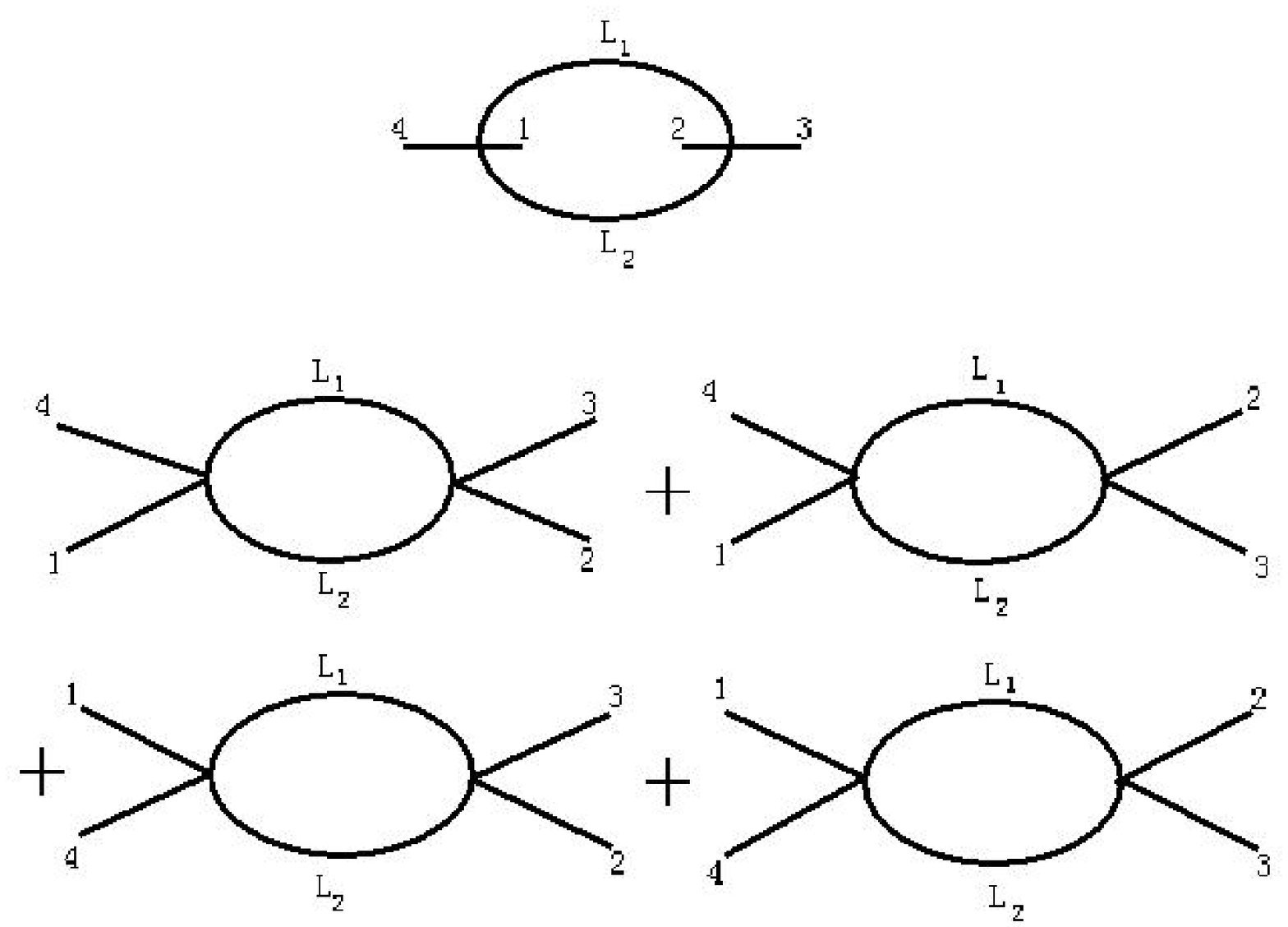}}
{\epsfxsize=2.50truein \epsfbox{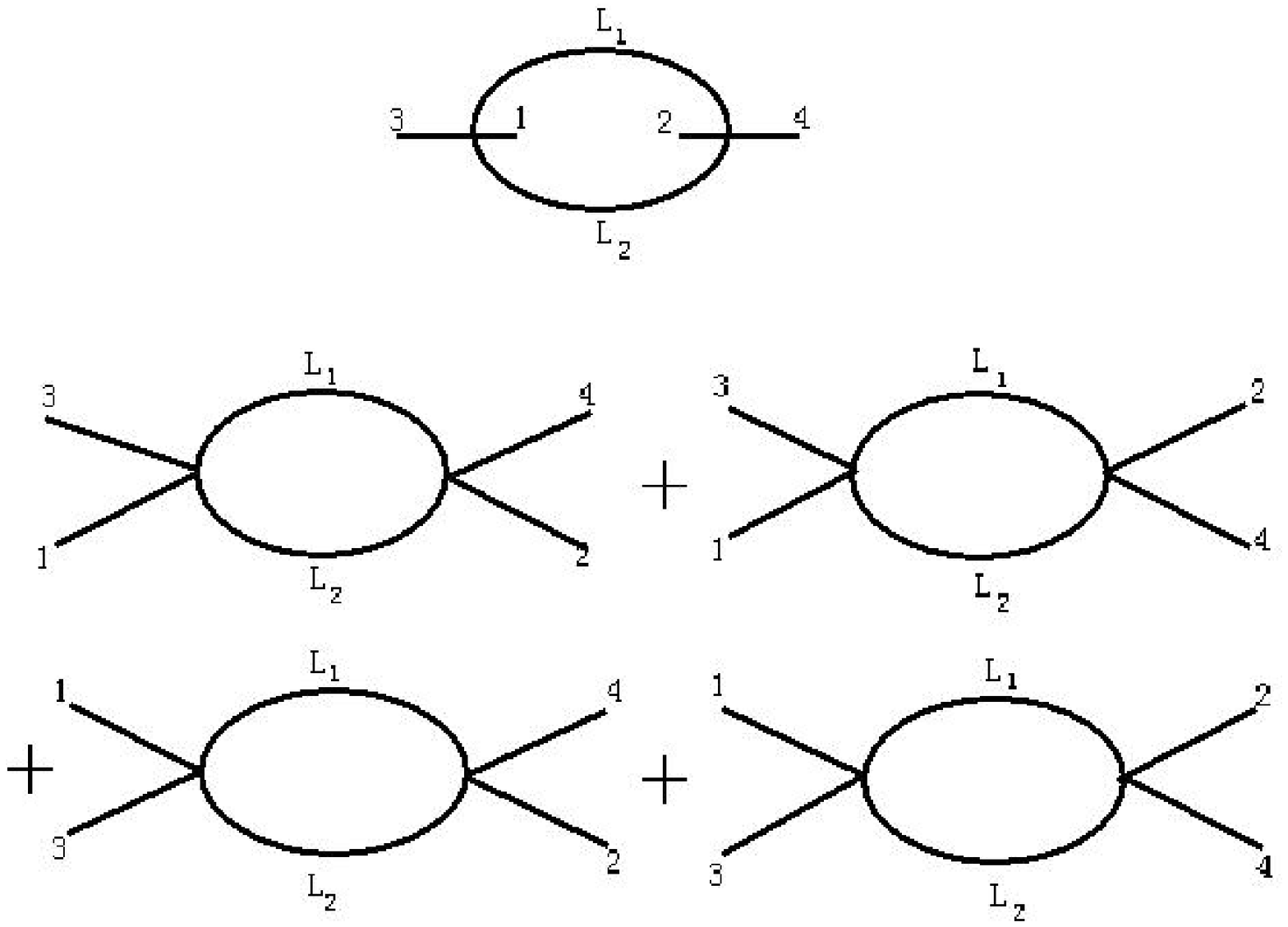}}
{\epsfxsize=2.50truein \epsfbox{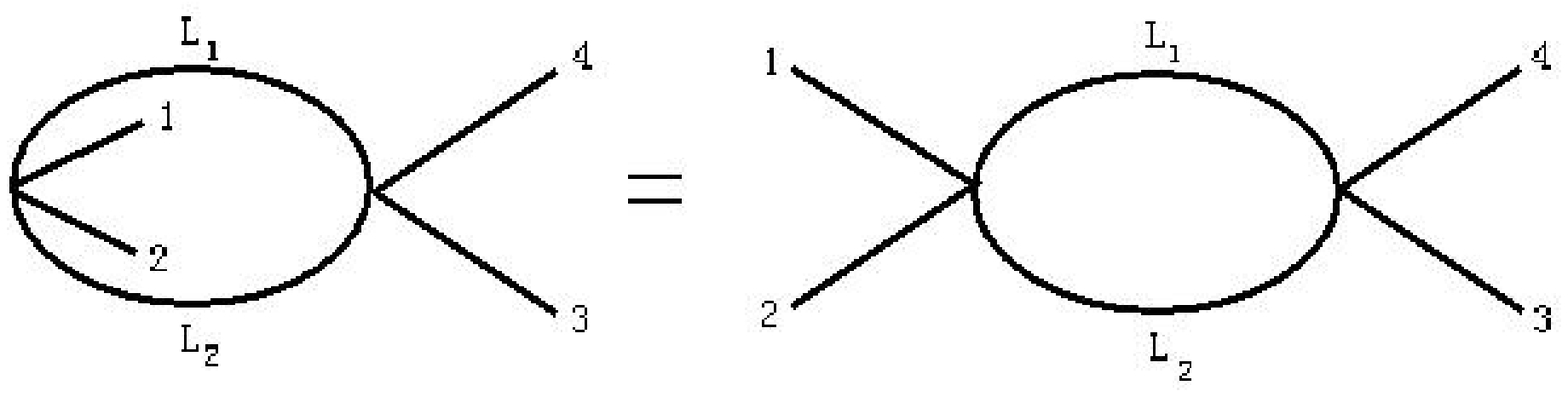}}
{\epsfxsize=2.50truein \epsfbox{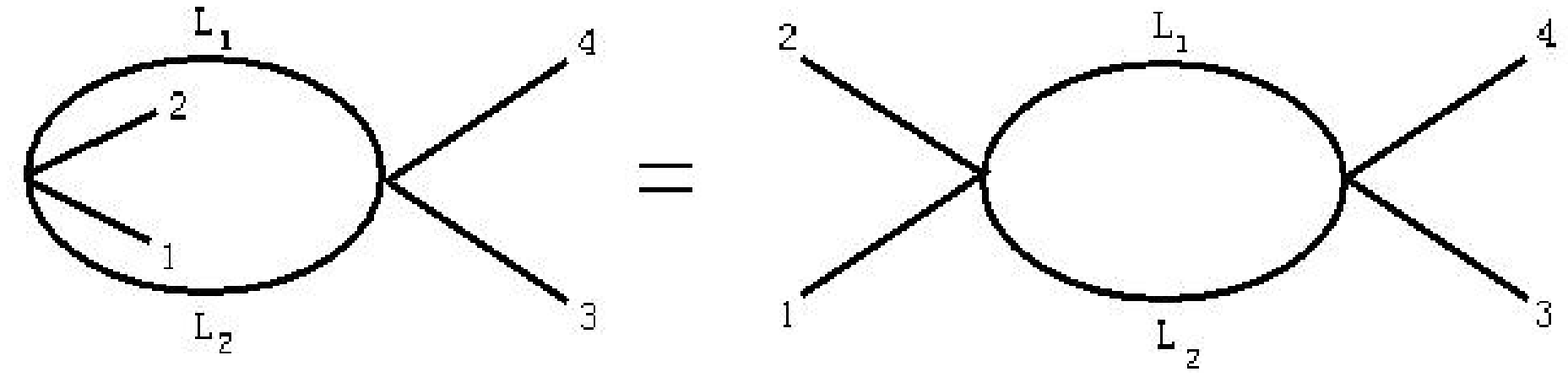}}
{\epsfxsize=2.50truein \epsfbox{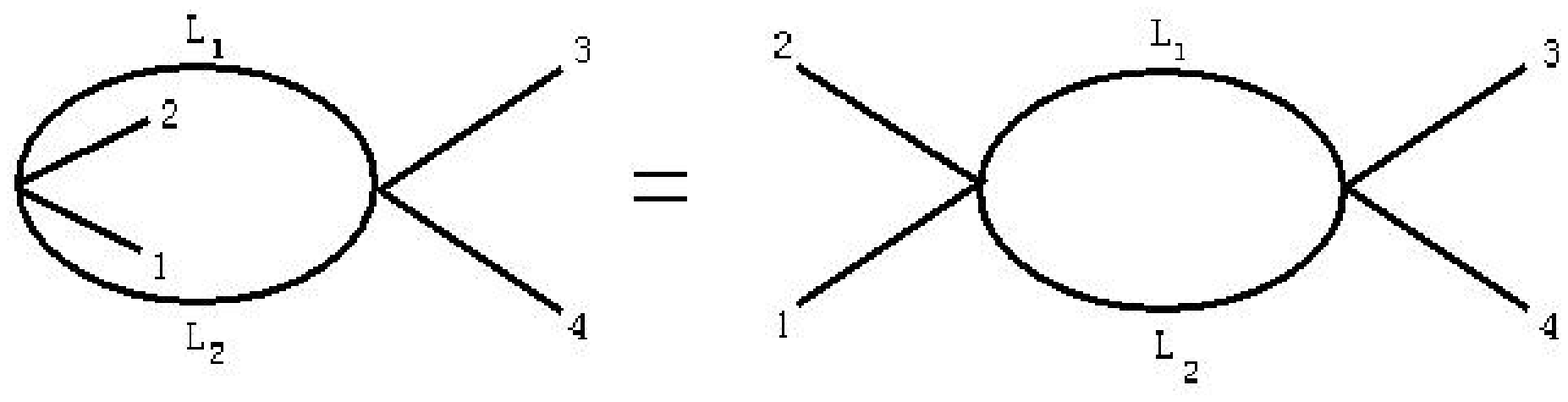}}
{\epsfxsize=2.50truein \epsfbox{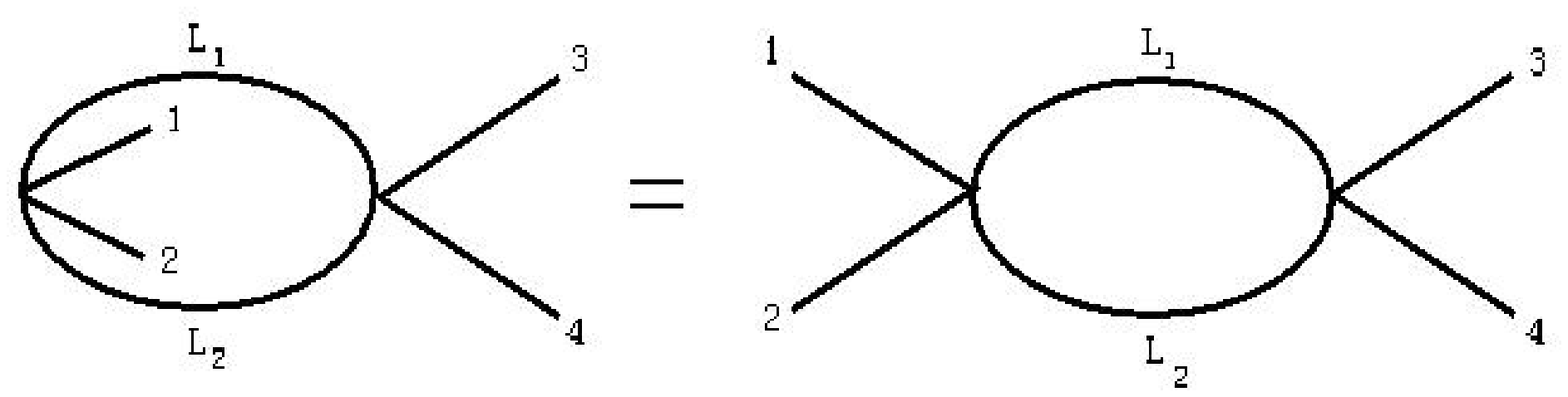}}
\end{center}
\caption{(b) Sub-leading one-loop MHV diagrams for the special case of four external particles.}
\end{figure}

We then turn to the sub-leading terms. 
For illustration, we take the terms with the color factor
$ {\rm Tr} \left( T^{a_1} T^{a_2} \right) {\rm Tr} \left(T^{a_3} T^{a_4} \right)$.
Terms with other color factors can be obtained 
by simple permutation of color indices and corresponding helicities and momenta.
By using Eq (\ref{Anc_field}), we can construct $A_{4;3}(1,2;3,4)$ in terms of the $A_{4;1}$ in Eq (\ref{eq3.5}).
This gives a similar integral as Eq (\ref{eq3.5}) but with the following integrand:
\be
F_0(1,2,3,4)+F_0(1,3,2,4)+F_0(3,1,2,4)+F_0(2,1,3,4)+F_0(2,3,1,4)+F_0(3,2,1,4)
\label{eq3.8}
\ee
Now the sub-leading MHV diagrams. There are six of them in total (Figure 3b), 
each gives a similar integral as Eq (\ref{eq3.5}) but with a different $h$ and related $P_L$: 
\be
\hat{h}_1(1,2;3,4) = \frac{\langle l_1,l_2 \rangle^4}
{\langle 4,l_1 \rangle\langle l_1,1 \rangle \langle 1,l_2 \rangle \langle l_2,4 \rangle
 \langle l_1,3 \rangle\langle 3,l_2 \rangle \langle l_2,2 \rangle \langle 2,l_1 \rangle}
\ee
with $P_L = \lambda_1 \tilde{\lambda}_1 + \lambda_4 \tilde{\lambda}_4$;
\be
\hat{h}_2(1,2;3,4) =\frac{\langle l_1,l_2 \rangle^4}
{\langle 3,l_1 \rangle\langle l_1,1 \rangle \langle 1,l_2 \rangle \langle l_2,3 \rangle
 \langle l_1,4 \rangle\langle 4,l_2 \rangle \langle l_2,2 \rangle\langle 2,l_1 \rangle}
\ee
with $P_L = \lambda_1 \tilde{\lambda}_1 + \lambda_3 \tilde{\lambda}_3$; and
\be
\hat{h}_3(1,2;3,4) =\frac{\langle l_1,l_2 \rangle^4}
{\langle l_1,1 \rangle \langle 1,2 \rangle \langle 2,l_2 \rangle \langle l_2,l_1 \rangle
 \langle l_1,4 \rangle \langle 4,3 \rangle \langle 3,l_2 \rangle \langle l_2,l_1 \rangle}
\ee
\be
\hat{h}_4(1,2;3,4) =\frac{\langle l_1,l_2 \rangle^4}
{\langle l_1,2 \rangle \langle 2,1 \rangle \langle 1,l_2 \rangle  \langle l_2,l_1 \rangle
 \langle l_1,4 \rangle \langle 4,3 \rangle \langle 3,l_2 \rangle \langle l_2,l_1 \rangle}
\ee
\be
\hat{h}_5(1,2;3,4)=\frac{\langle l_1,l_2 \rangle^4}
{\langle l_1,2 \rangle \langle 2,1 \rangle \langle 1,l_2 \rangle \langle l_2,l_1 \rangle
 \langle l_1,3 \rangle\langle 3,4 \rangle\langle 4,l_2 \rangle\langle l_2,l_1 \rangle}
\ee
\be
\hat{h}_6(1,2;3,4)=\frac{\langle l_1,l_2 \rangle^4}
{\langle l_1,1 \rangle\langle 1,2 \rangle\langle 2,l_2 \rangle \langle l_2,l_1 \rangle
 \langle l_1,3 \rangle\langle 3,4 \rangle\langle 4,l_2 \rangle\langle l_2,l_1 \rangle}
\ee
all with $P_L = \lambda_1 \tilde{\lambda}_1 + \lambda_2 \tilde{\lambda}_2$.
With the help of Schouten identity, we have
\ba
\hat{h}_1(1,2;3,4) = h_1(1,4,3,2)+h_1(1,4,2,3)+h_2(1,3,2,4)+h_2(1,2,3,4), \\
\hat{h}_2(1,2;3,4) = h_1(1,3,4,2)+h_1(1,3,2,4)+h_2(1,4,2,3)+h_2(1,2,4,3),  \nonumber \\
\hat{h}_3(1,2;3,4) = h_2(1,4,3,2),  \ \ \nonumber
\hat{h}_4(1,2;3,4) = h_1(1,2,4,3), \\ \nonumber 
\hat{h}_5(1,2;3,4) = h_1(1,2,3,4), \  \ 
\hat{h}_6(1,2;3,4) = h_2(1,3,4,2),
\ea
all graphically illustrated in Figure 3b. Putting all terms together, one gets
exactly the sum of F's in Eq (\ref{eq3.8}).

\begin{figure}[h]
\begin{center}
\leavevmode
{\epsfxsize=2.50truein \epsfbox{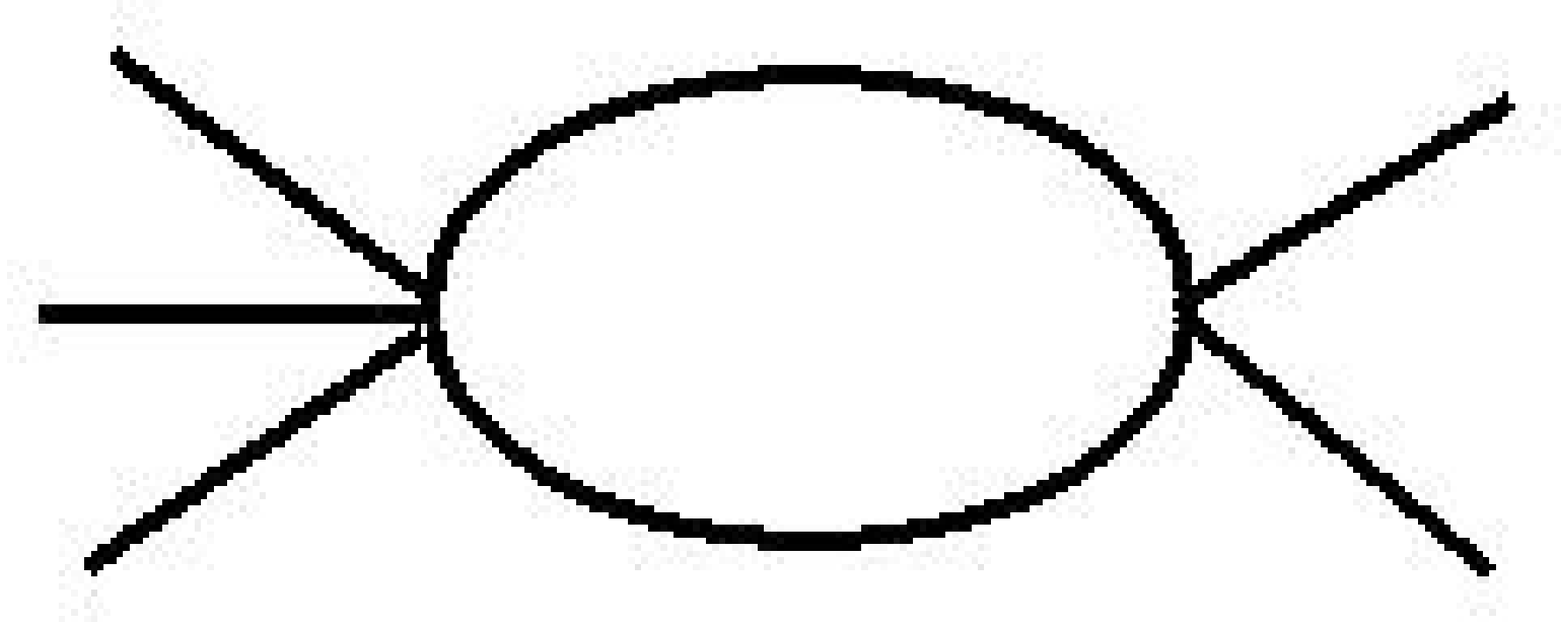}}
{\epsfxsize=6.00truein \epsfbox{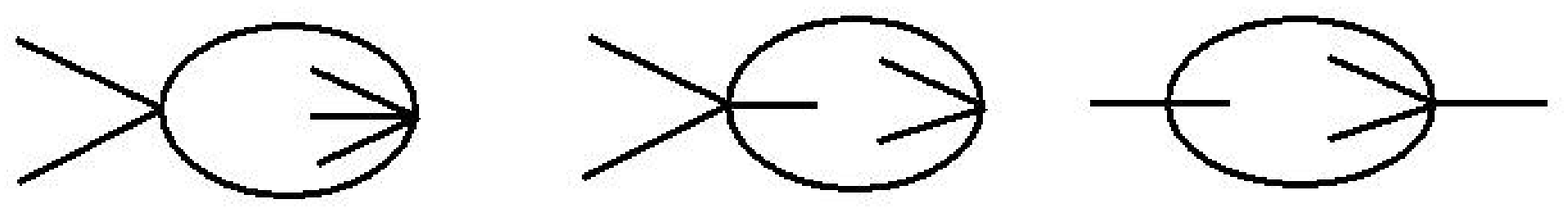}}
\end{center}
\caption{The topologies for leading (top) and sub-leading (bottom) one-loop MHV diagrams
         for the special case of five external particles.}
\end{figure}

For one-loop MHV diagrams of five external particles,
one vertex has four legs and the other has five legs. 
For a given color factor, ${\rm Tr} \left(T^{a_1} T^{a_2} T^{a_3} T^{a_4} T^{a_5} \right)$,
there is only one topology, as shown on the top of Figure 4, 
but five different MHV diagrams. 
These five diagrams combine to give a total integrand $F_0(1,2,3,4,5)$ 
and $A_{5;1}(1,2,3,4,5)$ after internal momentum integration.

For the sub-leading term with a given color factor,
${\rm Tr} \left(T^{a_1} T^{a_2} \right) {\rm Tr} \left( T^{a_3} T^{a_4} T^{a_5} \right)$,
we can similarly construct $A_{5;3}$ out of $A_{5;1}$. This gives an integral with the following integrand,
\ba
F_0(2,1,3,4,5) + F_0(2,3,1,4,5) + F_0(2,3,4,1,5) + F_0(3,2,1,4,5) \nonumber \\
+ F_0(3,2,4,1,5) + F_0(3,4,2,1,5) + F_0(1,2,3,4,5) + F_0(1,3,2,4,5) \nonumber \\
+ F_0(1,3,4,2,5) + F_0(3,1,2,4,5) + F_0(3,1,4,2,5) + F_0(3,4,1,2,5) \label{eq3.16}
\ea
Now the sub-leading MHV diagrams.
There are three topologies, as shown at the bottom of Figure 4, but sixty different MHV diagrams.
Each can be re-arranged as sum of diagrams in the leading term, after proper permutation of indices.
Shown in Figure 5 is one of the re-arrangement.
After a straightforward but lengthy calculation, the sixty diagrams are shown to 
yields the same expression in Eq (\ref{eq3.16}).
So in both cases, sub-leading MHV diagrams do give the correct sub-leading amplitudes.

\begin{figure}[h]
\begin{center}
\leavevmode
{\epsfxsize=5.50truein \epsfbox{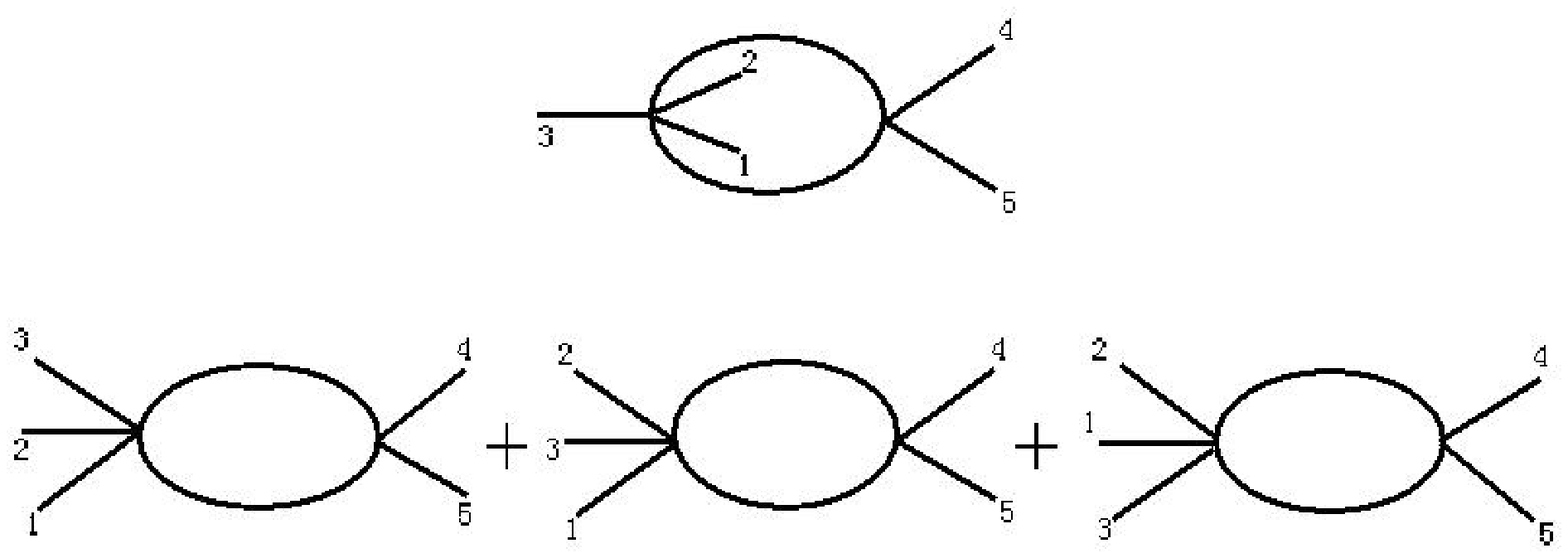}}
\end{center}
\caption{The decomposition of a sub-leading one-loop MHV diagram in terms of
leading one-loop MHV diagrams for one case of five external particles.}
\end{figure}

\section{General cases of arbitrary number of external particles}

In this section, we show that the CSW prescription gives
the correct full one-loop MHV amplitudes in the general case of
arbitrary number of external particles.
To fix notation, we first give the prescription to obtain the leading contribution.
For a given color factor, ${\rm Tr} \left(T^{a_1} \cdots T^{a_n} \right)$,
one gets all MHV diagrams from Figure 1, by distributing
external particles on the two vertices in all possible manners
but keeping the relative order according to the color indices.
Specifically, we first put $m=\{1, \cdots [n/2]\}$ external lines on the first vertex and the
rest on the second one.
As shown in the last section, the $m=1$ cases give a null result.
But they are kept here, as our results do not depend on the mentioned fact.
For each $m$, we put on the external lines clock-wise according to the color ordering to get a diagram,
and use cyclic permutation to get all different ones:
\be
A_{n;1}  = \sum_{m} 
\int {d^4 L_1 \ov L_1^2} {d^4 L_1 \ov L_1^2} 
\sum_{\sigma \in (\{\alpha\})} 
\delta^{(4)} (L_1+L_2+P_L) F(m,\sigma)
\ee
where 
\be
F(m,\sigma) =
 {\langle \sigma(m),\sigma(m+1)\rangle \langle \sigma(n),\sigma(1)\rangle
        \langle l_1, l_2 \rangle  \langle l_2, l_1 \rangle
              \ov  \langle l_2,\sigma(1) \rangle \langle \sigma(m), l_1 \rangle
                   \langle l_1,\sigma(m+1) \rangle \langle \sigma(n), l_2 \rangle}
      \prod_{i=1}^{n} {1 \ov \langle \sigma(i), \sigma(i+1)\rangle}
\ee
$P_L$ is the total momentum of all external particles on the left MHV vertex,
and $(\{\alpha\})$ is the set of all cyclic permutation of $n$ objects.
$A_{n;1}$ is identical to the ones obtained by conventional quantum field theory calculation \cite{bst}.

For a specific sub-leading term with the color factor
\be
{\rm Tr} \left( T^{a_1} \cdots T^{a_{c-1}} \right)
{\rm Tr} \left( T^{a_c} \cdots T^{a_n} \right)
\ee
we construct the sub-leading term $A_{n;c}(1,\cdots,c-1;c,\cdots, n)$
according to Eq (\ref{Anc_field}). A lengthy but straightforward calculation gives
\be
A_{n;c}  = \sum_m 
\int {d^4 L_1 \ov L_1^2} {d^4 L_2 \ov L_2^2} 
(-1)^{c-1} \sum_{\sigma \in COP^{''} \{\alpha\} \{\beta\}} 
\delta^{(4)} (L_1+L_2+P_L) F(m,\sigma)
\ee
where $\alpha_i \in \{\alpha\} \equiv \{c-1,\cdots,1\}$,
$\beta_i \in \{\beta\} \equiv \{c,c+1,\cdots,n\}$ 
and $COP^{''} \{\alpha\}\{\beta\}$ is the set of all permutations of
$\{1,\cdots,n\}$ that preserve the cyclic ordering of 
$\alpha_i$ within $\{\alpha\}$ and of $\beta_i$ within $\{\beta\}$, 
while allowing for all possible relative ordering of $\alpha_i$ with respect to $\beta_i$.

In the rest of the section, we will show that 
this $A_{n;c}(1,\cdots,c-1;c,\cdots, n)$ can also be obtained from all relevant sub-leading MHV diagrams. 
Explicitly, these sub-leading MHV diagrams are generated by distributing the first $c-1$ external particles 
inside the circle, and the rest $n-c+1$ external particles outside the circle,
but keeping the cyclic order inside and outside of the circle, respectively.

\begin{figure}
\begin{center}
\leavevmode
{\epsfxsize=5.0truein \epsfbox{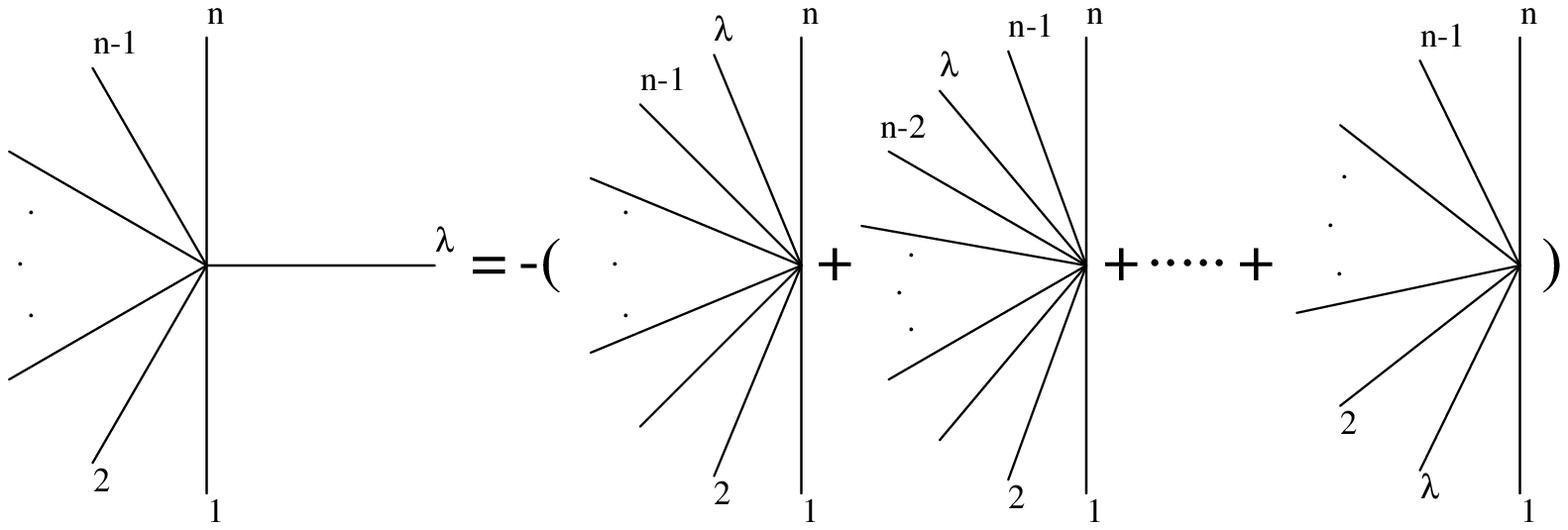}}
\end{center}
\caption{A graphic representation of Eq (4.5), one MHV amplitude is expressed in terms of other MHV amplitudes.
On the left-side of the equation, line $\lambda$ is to the right of lines $1$ and $n$.
On the right-side of the equation, line $\lambda$ is reflected to the left of lines $1$ and $n$,
and in all possible positions among the lines originally on the left.}
\end{figure}

Before starting, we note that the following relation between MHV amplitudes
\ba
\sum_{k=1}^{n-1} \left( \prod_{i=1}^{k-1}\frac{1}{\langle i,i+1\rangle} \right)
\frac{1}{\langle k ,\lambda \rangle \langle \lambda, k+1 \rangle}
\left( \prod_{i=k+1}^{n} \frac{1}{\langle i,i+1 \rangle} \right)
=- \frac{1}{\langle n,\lambda \rangle} \frac{1}{\langle \lambda,1 \rangle}
\prod_{i=1}^{n-1}\frac{1}{\langle i,i+1 \rangle}  \label{eq4.5} 
\ea
which can be proved in a straightforward manner by repeating use of the Schouten identity.
Shown in Figure 6 is a graphic representation of the identity.
By repeating use of this relation, one further gets
\ba
{1 \ov \langle \lambda, \alpha_{c-1} \rangle \langle \alpha_1, \eta \rangle
       \langle \eta, \beta_c \rangle \langle \beta_n, \lambda \rangle}
\left( \prod_{i=c-1}^2 {1 \ov \langle \alpha_i,\alpha_{i-1}\rangle} \right)
\left( \prod_{j=c}^{n-1}\frac{1}{\langle \beta_j,\beta_{j+1}\rangle} \right) 
 \nonumber \\
=
(-1)^{c-1}
{1 \ov \langle \lambda, \eta  \rangle}  
\sum_{\sigma \in COP^{'}\{\alpha\} \{\beta\}}
{1 \ov \langle \eta,\sigma(1) \rangle}
\left(  \prod_{i=1}^{n-1} {1 \ov  \langle \sigma(i),\sigma(i+1) \rangle} \right)
{1 \ov \langle \sigma(n), \lambda \rangle} 
\label{eq4.2}
\ea
where $\alpha_i \in \{\alpha\} \equiv \{c-1,\cdots,1\}$,
 $\beta_i \in \{\beta\} \equiv \{c,c+1,\cdots,n\}$ and $COP^{'}
 \{\alpha\}\{\beta\}$ is the set of all permutations of
 $\{1,\cdots,n\}$ that preserve the ordering of $\alpha_i$ within $\{\alpha\}$ and of $\beta_i$ within
 $\{\beta\}$, while allowing for all possible relative ordering
 of $\alpha_i$ with respect to $\beta_i$.
Figure 7 is a graphic representation of this identity.

\begin{figure}[h]
\begin{center}
\leavevmode
{\epsfxsize=5.0truein \epsfbox{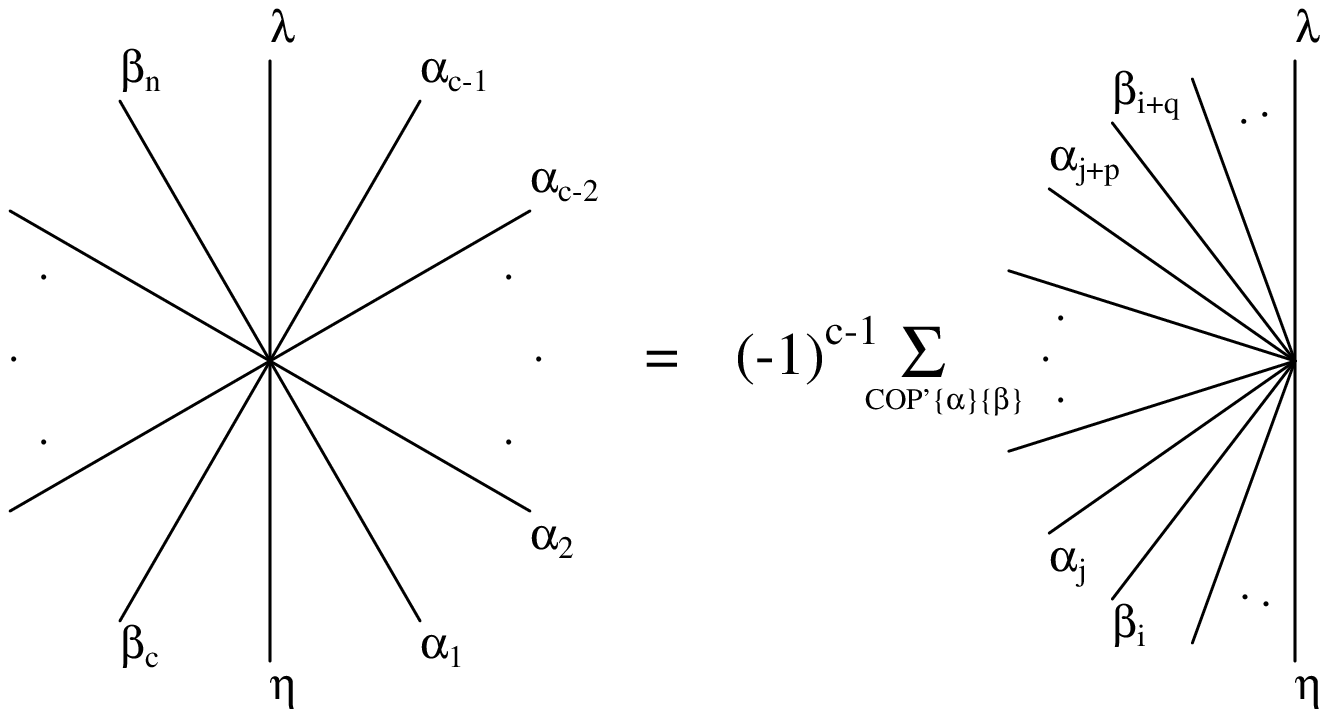}}
\end{center}
\caption{Graphic representation of Eq (4.6), another way to express one MHV amplitude in terms of
other MHV amplitudes.
On the left-side of the equation, all $\alpha$ lines are to the right of lines $\lambda$ and $\eta$.
On the right-side of the equation, all $\alpha$ lines are reflected to the left of lines $\lambda$ and $\eta$,
and in all possible positions among the lines originally on the left, 
while keeping the relative orders among themselves.}
\end{figure}

We now take a specific diagram with $m$ external lines on the first vertex,
among them, $j$ of which inside the circle and $i+1$ of which outside of the circle, 
$m=i+j+1$, as shown on the left of Figure 8.
Applying Eq(\ref{eq4.2}), we can make the following transformation for graphs on the left side
of Figure 8 to its right side:
\ba
\left( \prod_{r=c}^{c+i-1} \frac1{\langle \beta_r,\beta_{r+1} \rangle} \right)
 \frac1{\langle \beta_{c+i},l_1 \rangle \langle l_1,\alpha_1 \rangle}
 \left( \prod_{r=1}^{j-1} \frac1{\langle \alpha_r,\alpha_{r+1}\rangle} \right)
 \frac1{\langle \alpha_j,l_2\rangle \langle l_2,\beta_c \rangle} \nonumber \\
 \left( \prod_{r=j+1}^{c-2}\frac1{\langle \alpha_r,\alpha_{r+1}\rangle} \right)
 \frac1{\langle \alpha_{c-1},l_1\rangle\langle l_1,\beta_{c+i+1}\rangle}
 \left( \prod_{r=c+i+1}^{n-1} \frac1{\langle \beta_r,\beta_{r+1}\rangle} \right)
 \frac1{\langle \beta_n,l_2 \rangle\langle l_2,\alpha_{j+1}\rangle}
\nonumber \\
= 
(-1)^{c-1}
\left[ {1 \ov \langle \l_1,l_2 \rangle} 
\sum_{\sigma \in COP^{'} \{\alpha^{'}\}\{\beta^{'})\}} 
{1 \ov \langle l_2,\sigma(1) \rangle}
\left(  \prod_{i=1}^{m-1} {1 \ov  \langle \sigma(i),\sigma(i+1) \rangle} \right)
{1 \ov \langle \sigma(m), l_1 \rangle} \right]
\nonumber \\
\left[ {1 \ov \langle \l_2,l_1 \rangle} 
 \sum_{\sigma \in COP^{'} \{\alpha^{''}\}\{\beta^{''}\}} 
{1 \ov \langle l_1,\sigma(1) \rangle}
\left(  \prod_{i=1}^{n-m+1} {1 \ov  \langle \sigma(i),\sigma(i+1) \rangle} \right)
{1 \ov \langle \sigma(n-m), l_2 \rangle} \right] 
\label{eq4.7}
\ea
where $\alpha_i \in \{\alpha^{'}\} \equiv \{j,\cdots,1\} \ {\rm or} \ \{\alpha^{''}\} \equiv \{c-1,\cdots,j+1\}$, 
 $\beta_i \in \{\beta^{'}\} \equiv \{c,\cdots,c+i\} \ {\rm or} \ \in \{\beta^{''}\} \equiv \{c+i+1,\cdots,n\}$,
$COP^{'}  \{\alpha{'}\}\{\beta^{'}\}$ ($COP^{'}  \{\alpha{''}\}\{\beta^{''}\}$) 
is the set of all permutations of  $\{1,\cdots,m\}$ ($\{m+1,\cdots,n\}$) that preserve the ordering of 
$\alpha_i$ within $\{\alpha^{'}\}$ ($\{\alpha^{''}\}$)
 and of $\beta_i$ within  $\{\beta^{'}\}$ ($\{\beta^{''}\}$), 
while allowing for all possible relative ordering
 of $\alpha_i$ with respect to $\beta_i$.
Cyclically permuting all elements in $\{\alpha\}=\{\alpha^{'}\}\bigcup \{\alpha^{''}\}$ 
and $\{\beta\}=\{\beta^{'}\} \bigcup \{\beta^{''}\}$, and take the sum for all cases of $i$, $j$ and $m$, we get
\be
\sum_m (-1)^{c-1} \sum_{\sigma \in COP^{''}\{\alpha\} \{\beta\}} \delta^{(4)} (L_1+L_2+P_L) F(m,\sigma)
\ee 
which is nothing but the integrand in Eq (4.4).

Thus we have proved our main assertion that sub-leading amplitudes $A_{n;c}$ 
obtained from the sub-leading MHV diagrams
are related to the leading $N_c$ amplitudes $A_{n;1}$ 
in the same way as those obtained from conventional field theory calculations.
The CSW prescription gives the correct prediction for MHV amplitudes to one loop 
without taking the large $N_c$ limit.
\begin{figure}[h]
\begin{center}
\leavevmode
{\epsfxsize=6.0truein \epsfbox{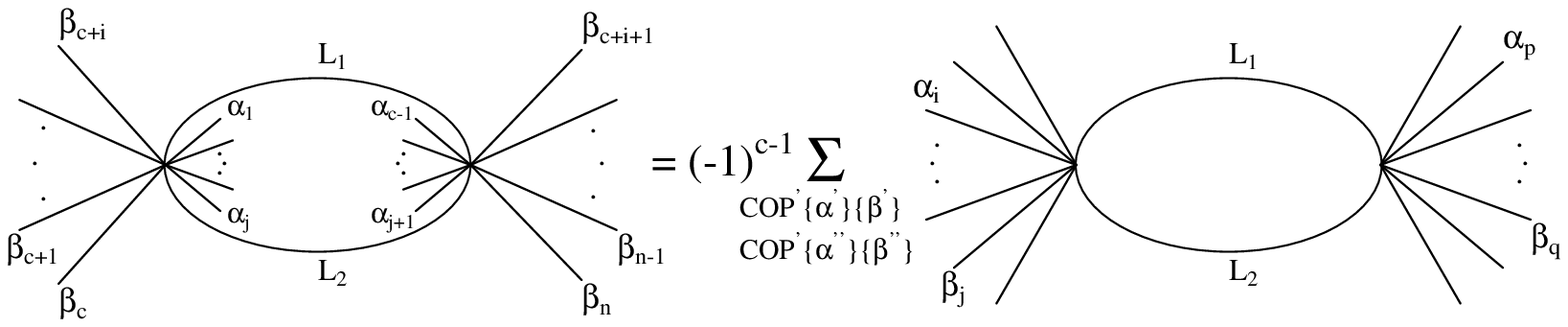}}
\end{center}
\caption{Graphic representation of Eq (4.7), 
a one-loop sub-leading MHV diagram is expressed in terms of a set of one-loop leading MHV diagrams.
All the external lines inside the circle are reflected to outside, in the manner prescribed in 
Eq (4.6) and Figure 7.}
\end{figure}

\section{Conclusion}
In this paper, we have analyzed the MHV amplitudes 
in general $N=4$ super Yang-Mills theories to full one-loop by using the CSW prescription. 
We reproduce the same relation
between leading $N_c$ amplitudes $A_{n;1}$ and sub-leading amplitudes $A_{n;c}$
as those obtained from conventional field theory calculations \cite{bern1}.
Combining with existing results \cite{bst}, this establishes the validity of the CSW approach
to one-loop in the calculation of MHV amplitudes, without taking the large $N_c$ limit.

It is tempting to speculate that the method is valid to non-MHV diagrams and to all orders,
though an understanding of the approach from the perspective of conventional
quantum field theory is still wanting.
From the viewpoint of quantum field theory,
the CSW prescription could be taken as an efficient organization principle of Feynman diagrams.
It is motivated by a string construction and
the string construction so far seems mainly to be related to the large $N_c$ limit.
The sub-leading results obtained in this paper indicate a wider applicability.
This calls for further inquiry of the rationale behind the prescription.
Parallel to the original motivation,
we believe that a construction of the prescription from conventional field theory
is feasible, as well as highly desirable, given the highly restrictive nature of $N=4$ SYM theories.
Such a construction may also change our perception of perturbative quantum field theories
with enormous symmetries.
In any case, it will be
interesting to extend the analysis to non-MHV diagrams as well as to higher order calculations.

\acknowledgments This work is supported in part by the National
Science Foundation of China.


\end{document}